\documentstyle[12pt,psfig,epsfig]{article}
\let\section=\subsection     \let\subsection=\subsubsection                

\begin{document}
\begin{center}
   {\large \bf S=I=0 PION PAIRS}\\[2mm]
   {\large \bf IN THE $A(\pi, 2\pi)X$ REACTION}\\[5mm]
   M.J.~VICENTE VACAS, E.~OSET
    \\[5mm]
   {\small \it Departamento de F\'{\i}sica Te\'{o}rica and IFIC,\\
   Centro Mixto Universidad de Valencia-CSIC,\\ 
   46100 Burjassot, Valencia, Spain. \\[8mm] }
\end{center}

\begin{abstract}\noindent
Recent experimental results show a large enhancement of the $2\pi$ emission
on the scalar isoscalar channel, which had been predicted by some theoretical 
estimates. We present here a detailed calculation of the   $A(\pi, 2\pi)X$
process incorporating a microscopic model for the elementary reaction in vacuum,
pion-pion final state interaction in the nuclear medium, and other nuclear
medium effects.
\end{abstract}

\section{Introduction}

The large, $A$ dependent  enhancement of the cross section at low invariant 
masses of the dipion system found in Refs. \cite{Cam:93,Bon:96,Bon:98,Bon:99}, 
could be a signal of very strong nuclear medium effects on correlated pion 
pairs in the $I=J=0$ ("$\sigma$") channel. This strength accumulation 
had been  predicted in Ref. \cite{Sch:88}, and similar results 
were obtained by Hatsuda et al. \cite{Hat:xx}, where the enhancement at low 
invariant masses of the spectral function in the  $\sigma$ channel appears 
as a consequence of the partial restoration of chiral symmetry in the nuclear 
medium.

 In the last few years, several non perturbative models have been developed, 
which describe very
successfully the $\pi\pi$ interaction in vacuum \cite{Jul:90,Oll:97}. 
When nuclear medium effects 
were included, some accumulation of strength was found close to the two pion 
threshold\cite{Rap:96,Aou:95,Chi:97}, which could be consistent with
the experimental results. However, a full calculation of the 
$A(\pi,2\pi)X$ process was needed in order to take into account all other 
nuclear effects and the detailed structure of the scattering amplitude.
 
  A first attempt was presented in ref. \cite{Rap:99}. In that work, a very 
simple model for the elementary $\pi N \to \pi\pi N$ amplitude was used, and
the most important medium effects were included. The results showed a clear peak
on the $M_{\pi\pi}$ distribution slightly above threshold, in good agreement
with experimental data for medium nuclei. However, the agreement was not as
satisfactory for deuterium, where the  $M_{\pi\pi}$ distribution was
overestimated at threshold.

In this paper we will present a  new study of the reaction, 
including a more realistic $\pi N \to \pi\pi N$ amplitude, which reproduces
very well the cross section on hydrogen and deuterium, and we shall also
consider some nuclear effects omitted previously, like the reduction of the 
incoming pion  flux due to absorption and quasielastic scattering, which 
modifies drastically the effective density at which the  reaction occurs.

\section{$\pi\pi$ scattering in the scalar isoscalar channel}

The $\pi\pi$ scattering amplitude is obtained solving the Bethe-Salpeter (BS) 
equation
\begin{equation}
T = {\mathcal{V}} +{\mathcal{V}}{\mathcal{G}} T.
\end{equation}
Fully detailed formulas and many technicalities can be found in Refs. 
\cite{Revc:99,Chi:97}. The $|\pi\pi ,I = 0>$ and $| K \bar{K}, I = 0>$ states 
are included in the coupled channels calculation. The potential ${\mathcal{V}}$ 
is obtained from the lowest order chiral lagrangians and $\mathcal{G}$ is the 
two meson propagator. A cutoff of 1 GeV is used to regularize the momentum
integral appearing in the calculation of  $\mathcal{G}$. The method guarantees
both unitarity and consistency with chiral perturbation theory at low energies.
  This theoretical  $\pi\pi$  scattering amplitude agrees well with experimental
phase shifts and inelasticities from threshold up to energies around 1.2 GeV, 
and therefore provides a good starting point for our analysis.

 To account for nuclear effects, the BS equation is modified by the
substitution of vacuum meson propagators by the medium ones. Namely, 
\begin{equation}
\tilde{\mathcal{G}} = i\int \frac{d^4 k}{(2 \pi)^4}
 \frac{1}{k^2 - m^2 -\Pi(k)} 
\frac{1}{(P - k)^2 - m^2-\Pi(P-k)},
\end{equation}
where $\Pi(k)$ is the meson selfenergy in  nuclear matter, which accounts for
the particle-hole and $\Delta$-hole excitations. 
 The resulting $\pi\pi$ scattering amplitude shows a strong dependence on the 
baryon density, as can be appreciated in Fig. 1. Whereas at high energies
(around 600 MeV) the imaginary part of $T_{\pi\pi}$ is reduced, there is 
a large enhancement around 300 MeV, where the CHAOS data show a well marked 
peak.
\begin{figure}[htb]
\centerline
{\psfig{figure=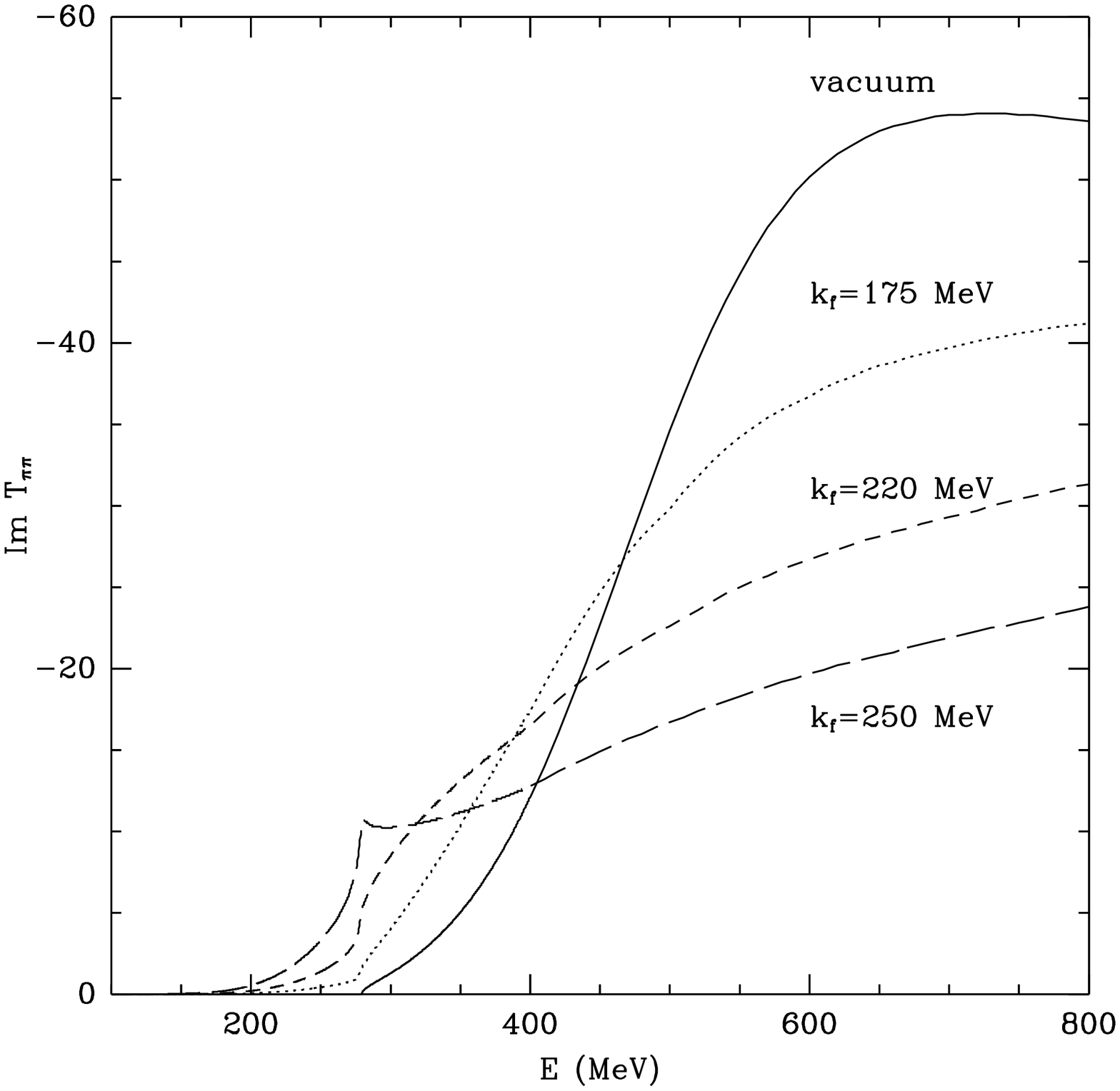,height=9.0cm,width=12.5cm,angle=0}}
\begin{small}
Fig.~1. {Im $T_{\pi\pi}$ in the S=I=0 scalar  channel as a function of 
the invariant mass of the pion pair and the nuclear density. Labels
correspond to the nucleons Fermi momentum.}\end{small}
\end{figure}
 Similar results have been  found using quite different approaches. See for 
instance   Ref.  \cite{Aou:95} and Fig. 3 of Ref. \cite{Sch:98}.
 
 Finally, let us mention that other isospin channels either have a very small 
contribution to the  $A(\pi,\pi\pi)X$ reaction at low energies, $(I=1)$, or 
show a  weak interaction between the pions $(I=2)$, thus, for them, we will not 
consider the interaction between the pions.

\section{Elementary $\pi N \rightarrow \pi \pi N$ reaction}

In order to be able to compare detailed effects on the differential cross
section a high quality description of the elementary cross section is clearly
required. Fortunately, such models are readily available
in the literature
\cite{Ose:85,Jae:92,Ber:95,Jen:97,Bol:97}. The model used in this paper follows
closely that of ref. \cite{Ose:85}, although with some improvements to accommodate 
it to new resonance data in the PDG book \cite{Cas:98}, and to properly include the
final state $\pi\pi$ interaction in the scalar isoscalar channel. The
Lagrangians and coupling constants used can be found in the appendix of 
ref. \cite{Revc:99}.

Although the model is quite complex, and includes many mechanisms, it has no 
free parameters. Some coupling constants, related to the Roper resonance, have 
an uncertainty band associated to the uncertainties quoted in the PDG book. In 
those cases we have always taken the central values.

 The results agree  well with the experimental data for total and
differential cross sections for all isospin channels\cite{Revc:99}, 
including of course the two-pions invariant mass distributions measured 
by CHAOS.

\section{$\pi A \rightarrow \pi \pi X$ reaction}
 
Many different nuclear effects modify the pion production cross sections. First,
the initial and final pions undergo a strong distortion. This is implemented
in the calculation following the methods of refs. \cite{Ose:86,Abs:94}.
The incoming pion flux is reduced by absorption and quasielastic scattering.
Both are very large because we are close to the $\Delta$ resonance  peak.
The pions scattered quasielastically are simply removed because they loose 
energy, and thus the probability to participate in a pion production process is
drastically reduced. Distortion is less important for the final pions because of
their lower energy. Only absorption has been considered for them. Second,
the incoming pion collides with a nucleon which is moving in a Fermi sea, and the
emitted nucleon  is Pauli blocked, therefore only momenta above certain value 
are allowed to contribute, and this is implemented by means of a local density
approximation. Third, the intermediate resonances 
($\Delta$'s, $N^*$'s) also see their properties modified by the medium. Also, 
new reaction mechanisms like meson exchange currents, could play some role, 
although it has been shown in Ref.  \cite{Bon:98} that the reaction is 
essentially quasifree, and these possible mechanisms are not included in our
calculation. 
Finally, the pion-pion final state interaction in the nuclear medium is
considered.  We select the part of the amplitude in which the two final 
pions are in the scalar isoscalar channel, and then we modify this part by 
incorporating the nuclear medium $\pi\pi$ interaction \cite{Revc:99}.

\section{Results and discussion}

As shown in Fig.~2, we find that our model describes fairly well 
the $\pi^+\to\pi^+\pi^+$ reaction both in deuterium and in heavier nuclei. 
Details and comments on normalization can be found in ref. \cite{Revc:99}.  
This gives us much confidence on our treatment of the nuclear medium effects. 
Note  that they are the same 
for this and the $\pi^+\pi^-$ channel except for the the two-pions final state
interaction, which is pure isospin 2 in $\pi^+\pi^+$, and mostly isospin 0 
in the $\pi^+\pi^-$ case.
\begin{figure}[htb]
\centerline
{\psfig{figure=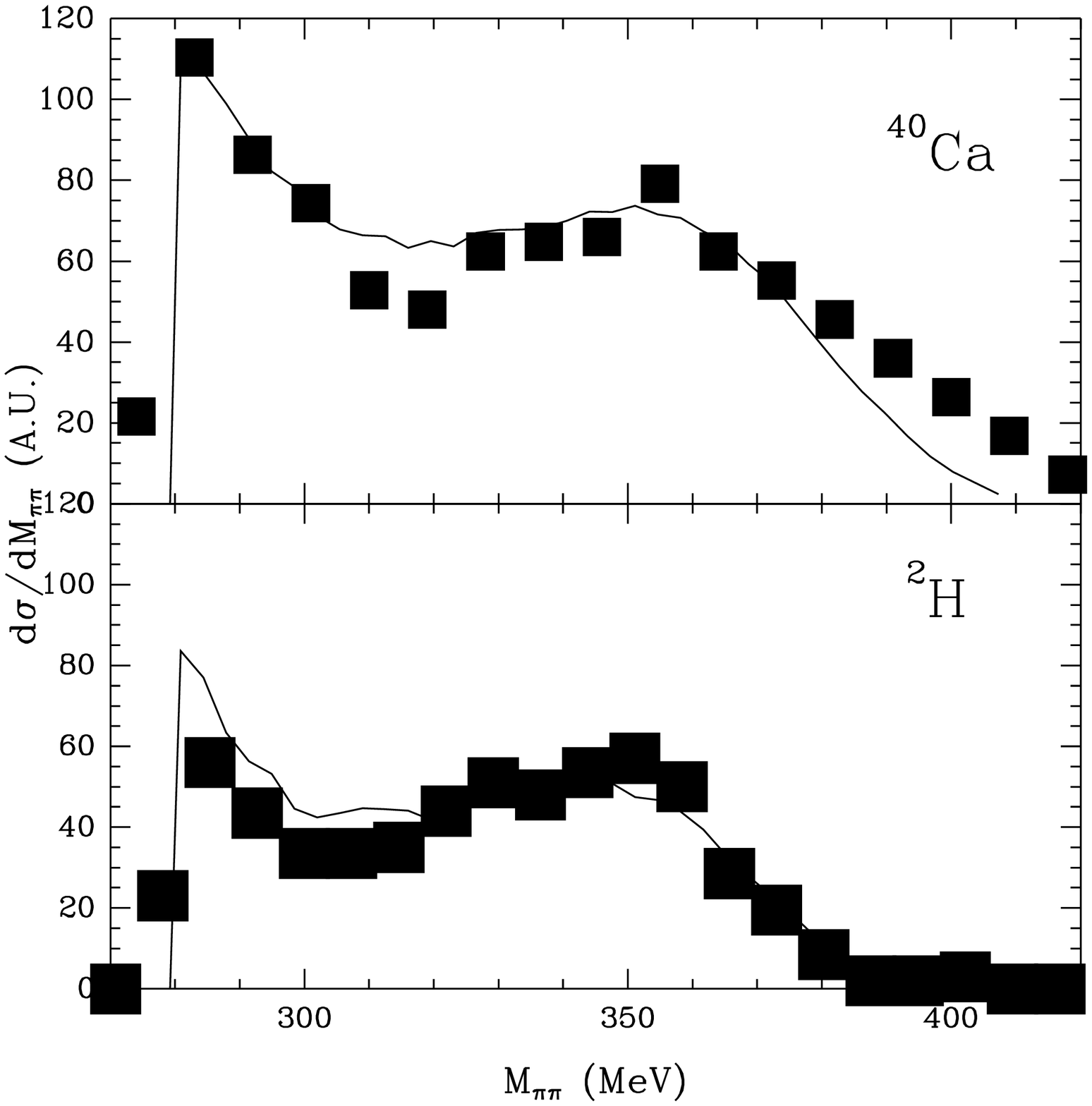,height=10.5cm,width=12.5cm,angle=0}}
\begin{small}
Fig.~2. {Two pion invariant mass distributions in the 
$\pi^++Ca\to \pi^+\pi^+ X$ (upper box), and $\pi^++\,^2H\to \pi^+\pi^+ X$ (lower box)
reactions. Experimental points are from  ref.   
\protect\cite{Bon:96}.}\end{small}
\end{figure}
\begin{figure}[htb]
\centerline
{\psfig{figure=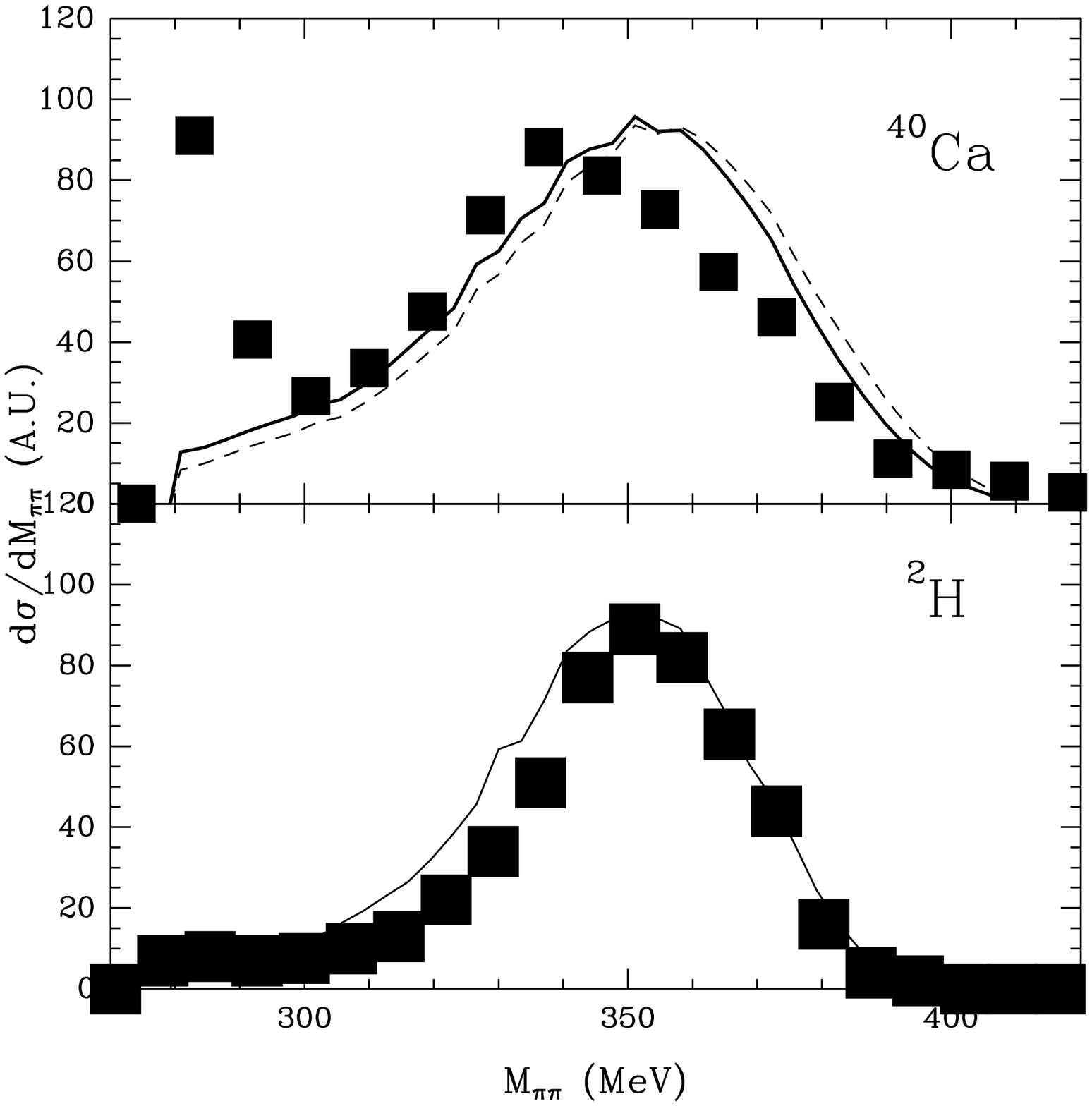,height=10.5cm,width=12.5cm,angle=0}}
\begin{small}
Fig.~3. {Two pion invariant mass distributions in the 
$\pi^-+Ca\to \pi^+\pi^- X$ (upper box), and $\pi^-+\,^2H\to \pi^+\pi^- X$ (lower box)
reactions. Solid line, full calculation; dashed line, no medium effects in the
FSI of the two pions. Experimental points are from  ref.   
\protect\cite{Bon:96}.}\end{small}
\end{figure}
However, although the  $\pi^-\to\pi^+\pi^-$ reaction is well
reproduced in deuterium, the model fails for this channel in
heavier nuclei (see Fig.~3). Furthermore, we find the effect of in-medium 
final state interaction of the pions to be rather small. Similar results are 
obtained for heavier nuclei.

The main reason for the small enhancement found is the  very small effective
density (see Fig.~4) at which the pion production process occurs. As explained 
before, the initial pion has a large probability of being
absorbed or quasielastically scattered. As a consequence, the flux reaching 
the center of the nucleus is  small and the reaction occurs mainly at the 
surface. An estimation of the average density gives $\rho_{av}=\rho_0/4$,
considerably lower than those used in Ref. \cite{Rap:99}.
\begin{figure}[htb]
\centerline
{\psfig{figure=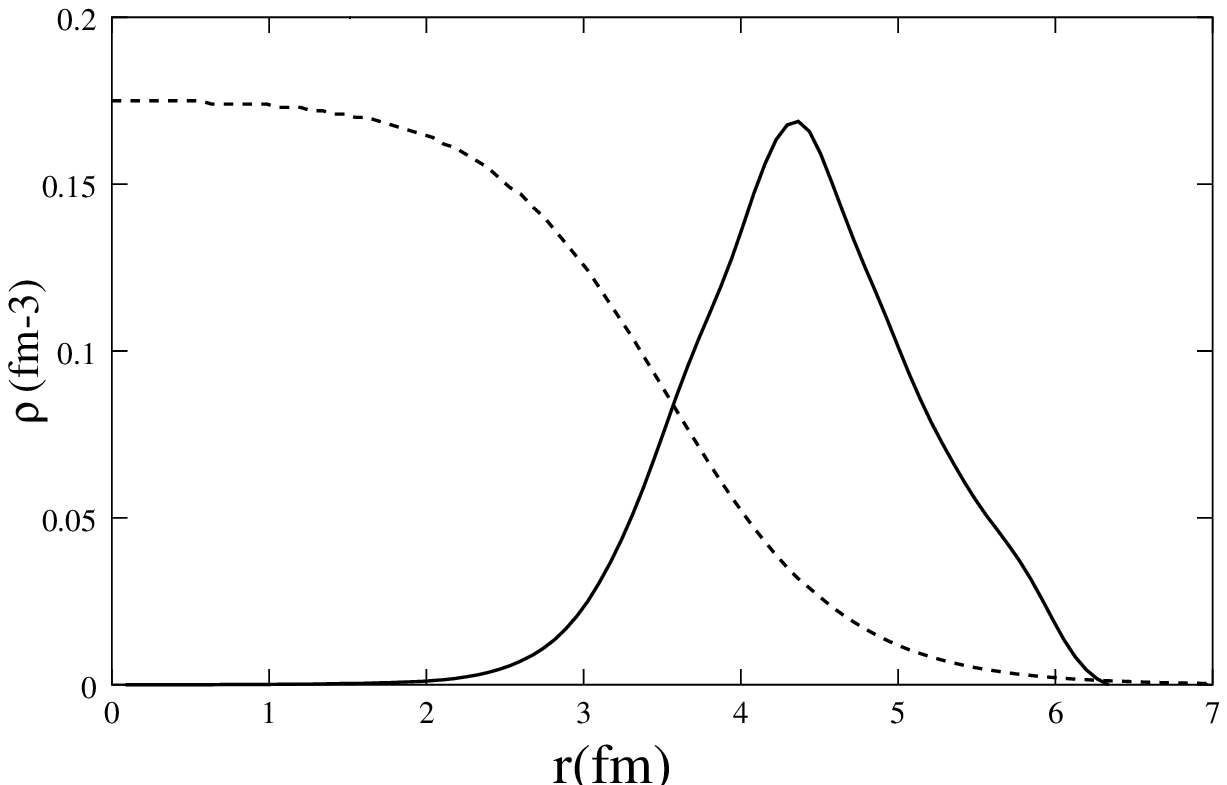,height=8.0cm,width=12.5cm,angle=0}}
\begin{small}
Fig.~4. {Solid line: Profile function showing the probability of a pion 
production event as a function of the radius for Calcium (arbitrary units). 
Dashed line: nuclear density.}\end{small}
\end{figure}
Imposing a high fixed average density we get a much larger although yet 
insufficient enhancement, and that at the price of destroying the nice agreement
found for the $\pi^+\pi^+$ case, due to the too large Fermi motion of the
nucleons.
 Better agreement with the nuclear data can be
reached by selecting a simplified version of the model used for the elementary
$\pi N \rightarrow \pi \pi N$ reaction, that overestimates the low mass region
for the deuteron case. 

There are several possibilities which could explain the large discrepancy 
between data and our model. One could be a much stronger pion-pion interaction
in the medium in the scalar isoscalar channels, and some recent works point into
that direction\cite{Cha:00}. On the other hand, more trivial effects could be
playing an important role. Probably, apart from concentrating on the 
peak appearing in medium and heavy nuclei, one should look more carefully
to the very low values at low invariant masses of the cross section in deuteron
and hydrogen. According to our model, this is due to destructive interference 
between large pieces of the amplitude. If some of these pieces are substantially
modified in nuclei,  the interference could disappear, and the spectral 
function would have some additional strength close to threshold.

 Some of these questions could be answered soon. There are new experimental
data being analyzed, which have measured a wider phase space than CHAOS, and
also other CHAOS measurements studying the energy dependence (meaning the 
incoming pion energy) of the peak. In our model, this is important because 
the interference effects are smaller at lower energies.

 Finally, we think that lepton induced reactions, which are free from the 
initial state interaction and would allow the pion production to happen at 
higher densities could be a better probe.
In particular, $(\gamma,\pi\pi)$ is currently under theoretical investigation.

\bigskip
\noindent
{\bf Acknowledgements}

\noindent
 This work has been partially supported by DGYCIT contract no. PB-96-0753.

\end{document}